\documentclass[a4paper, 12pt]{article}

\usepackage{amsmath, amssymb, amsthm}%, wasysym}
\usepackage[english]{babel}
\usepackage[T1]{fontenc}
\newcommand{\be}{\begin{equation}}
\newcommand{\ee}{\end{equation}}
\newcommand{\bd}{\begin{displaymath}}
\newcommand{\ed}{\end{displaymath}}
\newcommand{\ba}{\begin{eqnarray}}
\newcommand{\ea}{\end{eqnarray}}

\def\R{{I \!\! R}}

\def\v12{(v-w)}

\def\({\left(}
\def\){\right)}

\def\bgr#1\egr{{\allowdisplaybreaks\begin{gather}#1\end{gather}}}
\def\bma#1\ema{{\allowdisplaybreaks\begin{align}#1\end{align}}}

\def\oplem#1{\begin{lemma}\, {\rm #1}\, \it }
\def\cllem{\end{lemma}\rm \par }
\def\opthm#1{\begin{theorem}\, {\rm #1}\, \it }
\def\clthm{\end{theorem}\rm \par }

\def\R{\mathbb{R}}

\newcommand{\fer}[1]{(\ref{#1})}
\newcommand{\bq}{\begin{equation}}
\newcommand{\eq}{\end{equation}}
\def\bqa{\begin{eqnarray}}
\def\eqa{\end{eqnarray}}
\def\bd{\begin{displaymath}}
\def\ed{\end{displaymath}}

\theoremstyle{remark}

\theoremstyle{definition}

\begin{document}

\title{\Large An information-theoretic proof of Nash's inequality}
\author{Giuseppe Toscani \thanks{Department of Mathematics,
University of Pavia, via Ferrata 1, 27100 Pavia, Italy.
\texttt{giuseppe.toscani@unipv.it} }}

\maketitle

\begin{abstract}
We show that an information-theoretic property of Shannon's entropy
power, known as \emph{concavity of entropy power} \cite{Cos}, can be
fruitfully employed to prove inequalities in sharp form. In
particular, the \emph{concavity of entropy power} implies the
logarithmic Sobolev inequality, and the Nash's inequality with the
sharp constant.
\end{abstract}

\section{Introduction}

In information theory, inequalities constitute a powerful tool to solve
communication theoretic problems. Due to its wide range of application,
Shannon's entropy is at the basis of many of these inequalities
\cite{CTh}. Some deeper inequalities were developed by Shannon himself in
its pioneering 1948 paper \cite{Sha}. Among other facts, Shannon stated
the entropy power inequality in order to bound the capacity of
non-Gaussian additive noise channels.

In its original version, Shannon's entropy power inequality gives a
lower bound on Shannon's entropy functional of the sum of
independent random variables $X, Y$ with densities
 \be\label{entr}
\exp\left(\frac 2n H(X+Y)\right) \ge \exp\left(\frac 2n H(X)\right)+
\exp \left( \frac 2n H(Y)\right), \quad n \ge 1,
 \ee
with equality if $X$ and $Y$ are Gaussian random variables. In
inequality \fer{entr}, Shannon's entropy of a random variable $X$
with density is defined as
 \be\label{sh}
H(X) = H(f) = - \int_{\R^n} f(v) \log f(v)\, dv.
 \ee
Note that Shannon's entropy coincides with Boltzmann's $H$-functional up
to a change of sign \cite{Cer}.  The entropy-power
 \be\label{pow}
 N(X) = N(f) = \exp\left(\frac 2n H(X)\right)
 \ee
(variance of a Gaussian random variable with the same Shannon's
entropy functional) is maximum and equal to the variance when the
random variable is Gaussian, and thus, the essence of \fer{entr} is
that the sum of independent random variables tends to be \emph{more
Gaussian} than one or both of the individual components.

An interesting property of the entropy power has been discovered in 1985
by Costa \cite{Cos}. Let $f(v,t)$ denote the solution to the Cauchy
problem for the heat equation
 \be\label{heat}
\frac{\partial f (v,t)}{\partial t} =  \Delta f(v,t),
 \ee
posed in the whole space \cite{Cra}, corresponding to the initial value
$f(v)$, which we assume to be a probability density function.  Note that
for $t \ge 0$, the solution to the heat equation \fer{heat} can be written
as $f(v,t)= f*M_{2t}(v)$, where as usual $*$ denotes convolution, and
$M_t(v)$ is the Gaussian density in $\R^n$ of variance $n t$
 \be\label{max}
M_t(v) = \frac 1{(2\pi t)^{n/2}}\exp\left(\frac{|v|^2}{2t}\right).
 \ee
Costa \cite{Cos} proved that, for any given probability density function
$f$ different from the Gaussian density, $N(f*M_{2t})$ is a concave
function of time
 \be\label{conc}
\frac{d^2}{dt^2}N(f*M_{2t}) \le 0.
 \ee
The \emph{concavity property of entropy power} outlines a new
property of Gaussian functions. Indeed, the entropy power of a
Gaussian function coincides with its variance, so that the entropy
power of the fundamental solution to the heat equation is a linear
function of time. This linearity is restricted to Gaussian
densities.

Later, the original proof of Costa \cite{Cos} has been simplified in
\cite{Dem,DCT}, by an argument based on the Blachman-Stam inequality
\cite{Bla}. More recently, a short and physically relevant proof has been
obtained by Villani \cite{Vil}, resorting to some old ideas of McKean
\cite{McK}. The proof of Villani establishes a deep link between the
concavity of entropy power and the logarithmic Sobolev inequality. It is
remarkable that the same ideas of McKean have been seminal for a new proof
of logarithmic Sobolev inequality published some years ago \cite{To1}.

The concavity of entropy power involves the solution to the heat equation.
This basic fact includes the concavity of entropy power in the set of
inequalities which, in alternative to other ways of proof,  can be derived
by means of the heat equation. Indeed, the linear diffusion equation
\cite{To2} represents a powerful instrument to obtain a number of
mathematical inequalities in sharp form.

This maybe not so well-known property goes back more or less to half a
century ago, when independently from each others, researchers from
information theory \cite{Sta, Bla}, kinetic theory \cite{McK}, and
probability theory \cite{Lin}, established a useful connection between
Boltzmann's $H$-functional and Fisher information exactly by means of the
solution to the heat equation.

In this note, we proceed along the same lines to show that the
concavity of entropy power (a property of the solution to the heat
equation) allows to prove as corollaries important inequalities,
like the logarithmic Sobolev inequality and Nash's inequality in
sharp form.

Connections between the logarithmic Sobolev inequality and Nash's
inequality in sharp form are well known.  Beckner \cite{Bec1,Bec2} used
the former to prove the latter inequality with a sharp constant, thus
obtaining Nash's inequality from an argument different from the argument
used by Carlen and Loss \cite{Car}. The best constant for Nash's
inequality was indeed calculated by Carlen and Loss, who observed that
this inequality is equivalent to the Poincar\'e inequality in a suitable
ball of $\R^n$.

The next Section is devoted to the proof of the concavity of entropy
power. We will be mainly concerned with the key ideas behind this proof,
as well as to the analogies between this proof and analogous ones based on
the solution to the heat equation. Section \ref{sec-sob} will be devoted
to show that the logarithmic Sobolev inequality is a direct consequence of
the concavity of entropy power, which in some cases allows to prove the
previous inequality with a remainder.

Last, in Section \ref{sec-nash} we will show how Nash's inequality with a
sharp constant follows from the concavity of entropy power. The proof is
very simple, and makes use only of elementary inequalities, as well as of
well-known properties of the logarithmic function.

\section{The concavity of entropy power}\label{sec-con}

The proof of concavity  requires to evaluate two time derivatives of the
entropy power, along the solution $f(v,t)$ to the heat equation. The first
derivative of the entropy power is easily evaluated resorting to the
so--called DeBruijn's identity
 \be\label{e-f}
 I(f(t)) = \frac d{dt} H(f(t)), \quad t >0,
 \ee
which connects Shannon's entropy functional with the Fisher
information of a random variable with density
 \be\label{fish}
 I(X)= I(f) = \int_{\R^n} \frac{|\nabla f(v)|^2}{f(v)} \, dv.
 \ee
Using identity \fer{e-f} we get
 \[
\frac d{dt} N(f(t)) = \frac 2n \exp\left\{\frac 2n H(f(t))\right\}
\frac d{dt} H(f(t)) = \frac 2n \exp\left\{\frac 2n H(f(t))\right\}
\, I(f(t)).
 \]
Hence
 \[
\frac{d^2}{dt^2} N(f(t)) = \frac 2n \frac d{dt} \left[
\exp\left\{\frac 2n H(f(t))\right\} \, I(f(t))\right].
 \]
Let us set
 \be\label{Costa}
\Upsilon_f(t) = \Upsilon_f\left(f(t)\right) = \exp\left\{\frac 2n
H(f(t))\right\} \, I(f(t)).
 \ee
Then, the concavity of entropy power can be rephrased as the decreasing in
time property of the functional $\Upsilon_f(t)$ along the solution to the
heat equation. If
 \be\label{mc}
- J(f(t))= \frac{dI(f(t))}{dt},
 \ee
denotes the derivative of Fisher information along the solution to
the heat equation, we obtain
 \[
\frac d{dt}\Upsilon_f(t) = \exp\left\{\frac 2n H(f(t))\right\}
\left(
 \frac{dI(f(t))}{dt} + \frac 2n I(f(t))^2 \right) =
 \]
 \[
\exp\left\{\frac 2n H(f(t))\right\} \left(
 -J(f(t)) + \frac 2n I(f(t))^2 \right) .
 \]
Hence, $\Upsilon_f(t)$ is non increasing if and only if
 \be\label{key}
J(f(t)) \ge \frac 2n I(f(t))^2.
 \ee
It is interesting to remark that, aiming in proving the old
conjecture that subsequent derivatives of Boltzmann's $H$-functional
alternate in sign, the functional $J(f(t))$ was first considered by
McKean \cite{McK}. In one dimension, inequality \fer{key} is
essentially due to him. Let us repeat his highlighting idea. In the
one dimensional case one has
 \[
 I(f) = \int_\R
\frac{f'(v)^2}{f(v)} \, dv,
 \]
while
 \be\label{i22}
J(f) = 2 \left( \int_\R \frac{f''(v)^2}{f(v)} \, dv - \frac 13
\int_\R \frac{f'(v)^4}{f(v)^3} \, dv  \right).
 \ee
McKean observed that $J(f)$ is positive. In fact, resorting to
integration by parts,  $J(f)$ can be rewritten as
 \be\label{222}
J(f) = 2  \int_\R \left( \frac{f''(v)}{f(v)} -
\frac{f'(v)^2}{f(v)^2} \right)^2 \, f(v) \, dv \ge 0 .
 \ee
Having this formula in mind, consider that, for any constant
$\lambda >0$
 \[
0 \le 2 \int_\R \left( \frac{f''(v)}{f(v)} - \frac{f'(v)^2}{f(v)^2}
+ \lambda \right)^2 \, f(v) \, dv =
 \]
 \[
 J(f) + 2 \lambda^2 + 4 \lambda \int_\R \left( f''(v) -
\frac{f'(v)^2}{f(v)} \right) \, dv =
 J(f) + 2 \lambda^2 - 4 \lambda I(f).
 \]
Choosing $\lambda = I(f)$ shows \fer{key} for $n =1$.

Note that equality in \fer{key} holds if and only if $f$ is a Gaussian
density. In fact, the condition
 \[
\frac{f''(v)}{f(v)} - \frac{f'(v)^2}{f(v)^2} + \lambda = 0,
 \]
can be rewritten as
 \[
 \frac{d^2}{dv^2}\log f(v) = -\lambda,
 \]
which corresponds to
 \be\label{fin}
 \log f(v) = -\lambda v^2 + b v + c.
 \ee
Joining condition \fer{fin} with the fact that $f(v)$ has to be a
probability density, we conclude.

The argument of McKean was used by Villani \cite{Vil} to obtain \fer{key}
for $n
>1$. In the general $n$-dimensional situation, Villani  proved
the formula
 \[
J(f) = 2 \sum_{i,j = 1}^n \int_{\R^n} \left[
\frac{\partial^2}{\partial v_i \partial v_j}\log f \right]^2 f \, dv
=
 \]
  \be\label{n2}
2 \sum_{i,j = 1}^n \int_{\R^n} \left[\frac 1f
\frac{\partial^2}{\partial v_i \partial v_j} - \frac 1{f^2}
\frac{\partial f}{\partial v_i} \frac{\partial f}{\partial v_j}
\right]^2 f \, dv .
 \ee
By means of \fer{n2}, the nonnegative quantity
 \[
A(\lambda) = \sum_{i,j = 1}^n \int_{\R^n} \left[\frac 1f
\frac{\partial^2}{\partial v_i \partial v_j} - \frac 1{f^2}
\frac{\partial f}{\partial v_i} \frac{\partial f}{\partial v_j}
+\lambda \delta_{ij} \right]^2 f \, dv,
 \]
with the choice $\lambda = I(f)/n$, allows to recover inequality \fer{key}
for $n>1$. This proves the concavity property of entropy power.

To show that the concavity of entropy power has significant consequences,
we need to outline a further property of the functional $\Upsilon_f(t)$
\cite{To2}. Given a function $g(v)\ge 0, v \in \R^n $, let us consider the
scaling (dilation)
 \be\label{scal}
 g(v) \to g_a(v) = {a^n} g\left( {a} v \right), \quad a >0,
 \ee
which preserves the total mass of the  function $g$. By direct inspection,
it is immediate to conclude that Shannon's entropy \fer{sh} is such that,
if the probability density $f_a$ is defined as in \fer{scal}
 \be\label{h-scale}
H(f_a) = H(f) - n \log a.
 \ee
Since Fisher's information \fer{fish} scales according to
 \be\label{f-scale}
I(f_a) = \int_{\R^n} \frac{|\nabla f_a(v)|^2}{f_a(v)} \, dv = a^2
\int_{\R^n} \frac{|\nabla f(v)|^2}{f(v)} \, dv = a^2 I(f),
 \ee
one concludes that the functional $\Upsilon_f(t)$ is invariant with
respect to the scaling \fer{scal} of the solution $f(v,t)$ of the heat
equation. Therefore, for any constant $a
>0$
 \be\label{np}
\Upsilon_f(f(t)) =\Upsilon_{f}\left(f_a(t)\right).
 \ee
Property \fer{np} allows to identify the long-time behavior of the
functional $\Upsilon_f(t)$. Unless the initial value $f(v)$ in the heat
equation is a Gaussian function, the functional $\Upsilon(t)$ is monotone
decreasing, and it will reach its eventual minimum value as time $t\to
\infty$. The computation of the limit value uses in a substantial way the
scaling invariance property. In fact, at each time $t>0$, the value of
$\Upsilon_f(t)$ does not change if we scale the argument $f(v,t)$
according to
\begin{equation}\label{FP}
 f(v,t) \to F(v,t) = \left(\sqrt{1+2t}\right)^n\, f(v\, \sqrt{1+2t}, t),
\end{equation}
which is such that the initial value $f(v)$ is left unchanged. On the
other hand, it is well-known that (cfr. for example \cite{CT})
 \be\label{limi}
\lim_{t\to \infty} F(v,t) = M_1(v)
 \ee
where, according to \fer{max} $M_1(v)$ is the Gaussian density in $\R^n$
of variance equal to $n$. Likewise, the limit value of $\Upsilon_f(t)$
does not change if we scale the limit Gaussian function according to
\fer{scal} in order to have a variance different from one. Therefore,
passing to the limit one obtains, for any $\sigma>0$, the inequality
 \be\label{b5}
\exp\left\{\frac 2n H(f)\right\} \, I(f) \ge  \exp\left\{\frac 2n H(
M_\sigma )\right\} \, I(M_\sigma).
 \ee

\section{The logarithmic Sobolev inequality }\label{sec-sob}

Let us assume that $f(v)$ in \fer{b5} is a probability density function,
so that $\|f\|_{L^1} =1$. In this case, for any given probability density
$f(v)$, and any $\sigma >0$ inequality \fer{b5} takes the form
 \be\label{b6}
 \frac{I(f)}{I( M_\sigma)} \ge  \exp\left\{- \frac 2n \left( H(f) - H(
M_\sigma )\right)\right\}.
 \ee
Since
 \[
I( M_\sigma ) = \frac n\sigma,
 \]
while
 \[
H( M_\sigma ) = \frac n2 \log 2\pi\sigma + \frac n2,
 \]
using that $e^{-x} \ge 1-x$ we obtain from \fer{b6}
 \be\label{sob1}
\int_{\R^n} f(v)\log f(v) \, dv + n + \frac n2 \log 2\pi\sigma \le \frac
\sigma{2} \int_{\R^n} \frac{|\nabla f(v)|^2}{f(v)} \, dv .
 \ee
Inequality \fer{sob1} is nothing but the logarithmic Sobolev
inequality by Gross \cite{Gro}, written in an equivalent form.

Consider now the case in which the probability density $f(v)$ of the
random variable $X$ is such that the second moment of $X$ is
bounded. Then, for any $\sigma$ such that
 \be\label{con1}
 \sigma \ge \frac 1n \int_{\R^n} |v|^2 f(v) \, dv,
 \ee
it holds
 \[
-H(f) + H(M_\sigma) = \int_{\R^n} f(v)\log f(v) \, dv - \int_{\R^n}
M_\sigma (v) \log M_\sigma (v) \, dv =
 \]
 \[
 \int_{\R^n} f(v) \log
\frac{f(v)}{M_\sigma(v)} \, dv + \frac 1{2\sigma}\int_{\R^n}|v|^2
\left( M_\sigma - f(v) \right) \, dv \ge  \int_{\R^n} f(v) \log
\frac{f(v)}{M_\sigma(v)} \, dv.
 \]
By the Csiszar-Kullback inequality \cite{Kul}
 \be\label{ck}
2 \int_{\R^n} f(v) \log \frac{f(v)}{M_\sigma(v)} \, dv \ge \|
f-M_\sigma \|_{L^1}^2.
 \ee
By expanding the right-hand side of inequality \fer{b6} up to the
second order, we end up with the inequality
 \be\label{imp}
\frac \sigma{2} \int_{\R^n} \frac{|\nabla f(v)|^2}{f(v)} \, dv -
\int_{\R^n} f(v)\log f(v) \, dv + n + \frac n2 \log 2\pi\sigma \ge
\frac{n^2}8 \| f-M_\sigma \|_{L^1}^4.
 \ee
The right-hand side of \fer{imp} constitutes an improvement of the
logarithmic Sobolev inequality, in that, at least when the density
function involved into inequality \fer{b6} has bounded second
moment, and it is different from a Gaussian density, it is possible
to quantify the positivity of the difference between the right and
left sides of \fer{b6} in terms of the distance of it from the
manifold of the Gaussian densities, with a precise estimate of this
distance in terms of the $L^1$-norm.

\section{Nash's inequality revisited}\label{sec-nash}

A second interesting consequence of the concavity of entropy power is the
proof of a reinforced version of Nash's inequality \cite{Nash}. To this
aim, note that the right-hand side of inequality \fer{b5}, thanks to the
scaling invariance property of $\Upsilon_f$, does not depend of $\sigma$.
Since for any given $\sigma>0$
 \[
I( M_\sigma ) = \frac n\sigma,
 \]
while
 \[
H( M_\sigma ) =  \frac n2 \log 2\pi\sigma + \frac n2.
 \]
The choice
 \be\label{ok}
 \sigma = \bar\sigma = (2\pi e)^{-1},
 \ee
gives
  \[
I( M_{\bar\sigma} ) = 2\pi e n,
 \]
and
 \[
H(M_{\bar\sigma} ) =  0.
 \]
Thus, substituting the value $\sigma = \bar\sigma$ in \fer{b5} we
obtain the inequality
  \be\label{b8}
\exp\left\{\frac 2n H(f)\right\} \, I(f) \ge 2\pi e n.
 \ee
 Inequality \fer{b8} is know under the name of \emph{Isoperimetric Inequality for Entropies} (cfr. \cite{DCT} for a
different proof).

The case in which $f(v)\ge 0$ is a density of mass different from $1$,
leads to a modified inequality. Let us set
 \[
\mu = \int_{\R^n} f(v) \, dv.
 \]
Then, the function $\phi(v) = f(v)/\mu$ is a probability density, which
satisfies \fer{b8}. Therefore
 \[
I(\mu\phi) = \mu \, I(\phi)\ge
 \mu \,  I(M_\sigma)\exp\left\{\frac 2n H(M_\sigma)\right\} \,
 \,\exp\left\{- \frac 2n H(\phi)\right\} =
 \]
 \[
\mu \,  I(M_\sigma)\exp\left\{\frac 2n ( H(M_\sigma)- \log \mu)\right\} \,
 \,\exp\left\{- \frac 2n ( H(\phi) - \log \mu)\right\} =
 \]
 \be\label{q1}
\mu \,  I(M_\sigma) \exp\left\{\frac 2n \frac 1\mu H(\mu M_\sigma)\right\}
\,
 \,\exp\left\{- \frac 2n \frac 1\mu H(\mu \phi)\right\}.
 \ee
In \fer{q1} we used the identity
 \[
 H(\mu\phi) = \mu H(\phi) - \mu \log \mu.
 \]
Setting now $\sigma = \bar\sigma$, as given by \fer{ok}, we conclude with
the inequality
 \be\label{gen}
I(f) \ge 2\pi e n\, \|f\|_{L^1} \exp\left\{- \frac 2{n \|f\|_{L^1}} \left[
H(f)- \|f\|_{L^1} \log \|f\|_{L^1} \right]\right\},
 \ee
which clearly holds for any integrable function $f(v) \ge 0$.

Given a probability density function $g(v)$, let us set $f(v) = g^2(v)$.
In this case
 \[
H(f) = H(g^2)=  - \int_{\R^n} g^2(v)\log g^2(v) \, dv = -2
\int_{\R^n} \left( g(v)\log g(v)\right) g(v) \, dv.
 \]
Since the function $h(r) = r \log r$ is convex, and $\|g\|_{L^1} =
1$,  Jensen's inequality implies
 \be\label{bel}
- H(g^2) \ge  2 \int_{\R^n} g^2(v) \, dv \log \int_{\R^n} g^2(v) \,
dv.
 \ee
Using \fer{bel} into \fer{gen} gives
  \be\label{b70}
I(g^2) \ge 2\pi e n \int_{\R^n} g^2(v)\, dv \cdot  \exp\left\{ \frac 2n
\log \int_{\R^n} g^2(v) \, dv
 \right\} = \left( \int_{\R^n} g^2(v) \, dv \right)^{1 + 2/n}.
 \ee
Using the identity
 \[
 I(g^2) = 4 \int_{\R^n} |\nabla g(v) |^2 \, dv
 \]
we obtain from \fer{b7} the classical Nash's inequality in sharp
form
 \be\label{nass}
 \left( \int_{\R^n} g^2(v) \, dv \right)^{1+ 2/n} \le   \frac 2{\pi e n} \int_{\R^n} |\nabla g(v) |^2 \, dv
 \ee
Inequality \fer{nass} clearly holds for all probability density
functions $g(v)$. Note that, if $\|g\|_{L^1} \not= 1$, \fer{nass}
implies
  \be\label{nass1}
\left( \int_{\R^n} g^2(v) \, dv \right)^{1+ 2/n} \le   \frac 2{\pi e
n} \left( \int_{\R^n}|g(v)\, dv \right)^{4/n} \int_{\R^n} |\nabla
g(v) |^2 \, dv.
 \ee
The constant $2/(\pi e n)$ in \fer{nass1} is sharp.

\section{Conclusions}

The concavity of entropy power is a property of Shannon's entropy which
has unexpected consequences in terms of functional inequalities. In this
paper we made explicit the links between this property and the logarithmic
Sobolev inequality by Gross \cite{Gro}, as well as Nash's inequality
\cite{Nash}. In both cases, the concavity of entropy power allows to
improve these inequalities. In the case of the logarithmic Sobolev
inequality, it is shown that, for densities with bounded second moment, it
is possible to give a precise estimate of the distance between the density
and the manifold of Gaussian functions, which are known to saturate the
inequality.  Also, the clearness of the physical idea, and the relative
simplicity of the underlying computations, are in favor of the
information-theoretic proof of these inequalities.
\bigskip \noindent

{\bf Acknowledgment:} This paper has been written within the activities of
the National Group of Mathematical Physics of INDAM (National Institute of
High Mathematics). The author acknowledge support by MIUR project
``Optimal mass transportation, geometrical and functional inequalities
with applications''.

\end{document}